\documentclass[amsmath,amssymb,aps,prl,  twocolumn,superscriptaddress,notitlepage]{revtex4-2}

\usepackage{graphicx}
\usepackage{bbm}
\usepackage{dcolumn}
\usepackage{xcolor}

\newcommand{\pmn}{ Pb$(\text{Mg}_{1/3}\text{Nb}_{2/3})\text{O}_3$ }

\begin{document}

\title{Intrinsic structure of relaxor ferroelectrics from first principles}

\author{Xinyu Xu}
\thanks{X.X. and K.C. contribute equally to this work}
\affiliation{Applied Mathematics and Computational Research Division, Lawrence Berkeley National Laboratory, Berkeley, CA 94720, USA}
\affiliation{Department of Materials Science and Engineering, Stanford University, Stanford, CA 94305, USA}

\author{Kehan Cai}
\thanks{X.X. and K.C. contribute equally to this work}
\affiliation {Department of Chemistry, Princeton University, Princeton, NJ 08544, USA}

\author{Yubai Shi}
\affiliation{Department of Materials Science and Engineering, National University of Singapore, Singapore}

\author{Peichen Zhong}
\affiliation{Department of Materials Science and Engineering, National University of Singapore, Singapore}

\author{Pinchen Xie}
\thanks{Contact Author: pinchenxie@lbl.gov}
\affiliation{Applied Mathematics and Computational Research Division, Lawrence Berkeley National Laboratory, Berkeley, CA 94720, USA}
\date{\today}

\begin{abstract}
  We develop FIRE-Swap, a first-principles framework for sampling intrinsic compositional structures in complex perovskites with machine-learning interatomic potentials (MLIPs). Using both dedicated and universal MLIPs, we study the relaxor lead magnesium niobate (PMN) and the solid solutions lead zirconate titanate (PZT) and lead strontium titanate (PST).
  Across MLIP models and exchange-correlation approximations, FIRE-Swap robustly predicts a rock-salt-like chemical order in PMN, which is absent in PZT and PST with the same mixing ratio, consistent with experiments. We further identify in PMN a distinct Nb-cluster morphology. Interconnected, non-coarsened polar nanoregions are found within Nb clusters, providing a mesoscale basis for understanding relaxor ferroelectricity.
\end{abstract}

\maketitle

Relaxor ferroelectrics, or relaxors, are glass-like crystals that respond strongly to electric fields over wide ranges of temperature and frequency~\cite{smolenskii1961ferroelectrics, burns1983glassy, burns1983crystalline, viehland1990freezing, viehland1991glassy,  levstik1998glassy, cowley2011relaxing, bokov2020recent}. Their hallmark is compositional disorder that disrupts the uniform ferroelectric distortion of conventional ferroelectrics~\cite{rabe2007physics}.

\begin{figure*}[tb]
  \centering
  \includegraphics[width=0.9\linewidth]{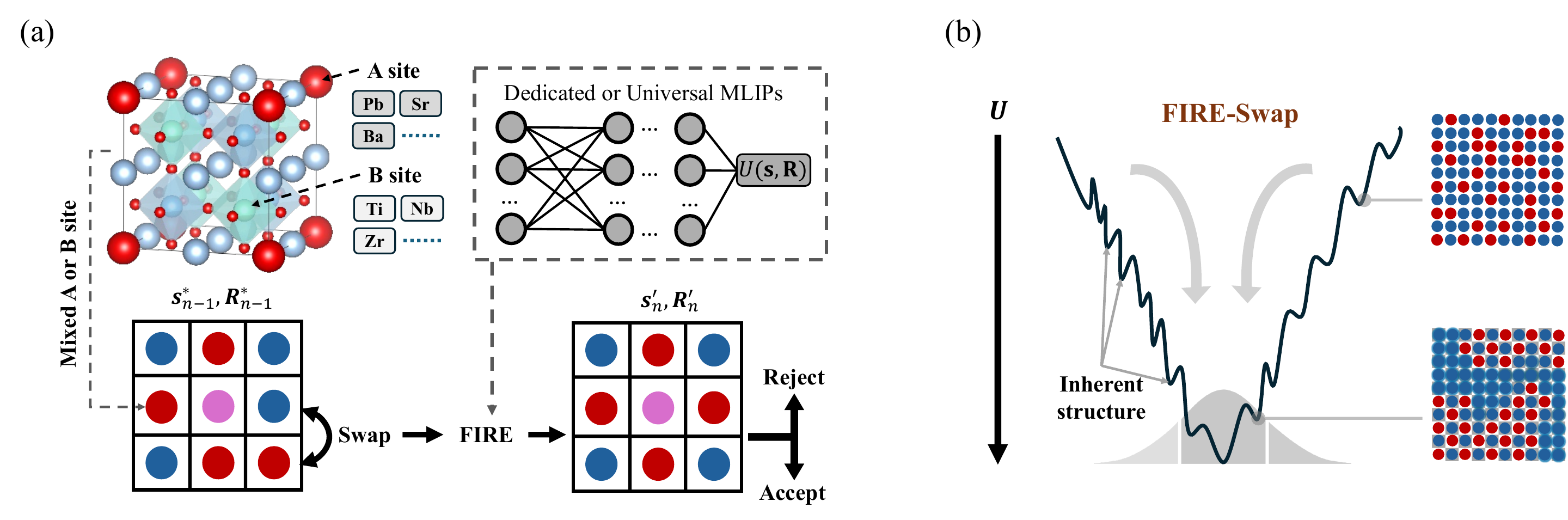}
  \caption{(a) Sketch of a $3\times 3\times 3$ supercell of a complex perovskite and one iteration of the FIRE-Swap algorithm. (b) Schematic representation of the PES as a funnel. A FIRE-Swap simulation traverses the energy landscape and equilibrates to the Boltzmann distribution of inherent structures, illustrated as the grey distribution at the bottom. The two CS illustrations to the right of the PES represent a fully disordered state and a state with partial chemical order.}
  \label{fig:workflow}
\end{figure*}

Many relaxors are complex perovskites based on the $\mathrm{ABO_3}$ structure, with mixed A- or B-site occupancy by multiple metallic species (see Fig.~\ref{fig:workflow}(a)). The resulting compositional disorder implies random local fields and frustrated couplings, which can be described within a spin-glass framework~\cite{sherrington2013bzt}, consistent with the phenomenological similarity between the permittivity peaks of relaxors and spin glasses~\cite{lecomte1983frequency, sherrington2013bzt}. However, spin-glass models assume homogeneous disorder and therefore miss intrinsic compositional structures (CS) in real materials, including short- and possibly long-range chemical order that can control macroscopic responses.
Some complex perovskites, such as \pmn, are heterovalent insulators in which charge, polar, and elastic degrees of freedom are strongly coupled, yielding a rugged energy landscape. This differs from metallic alloys, where screening and the absence of a polar lattice permit reduced lattice models~\cite{porter2009phase} with simple effective interactions (e.g., Ising-type Hamiltonians). These differences motivate a first-principles framework that treats compositional and geometric relaxations concurrently~\footnote{See Appendix C for further discussion}.

Conventional first-principles approaches remain limited. Density functional theory (DFT) calculations are constrained to small supercells. Molecular dynamics (MD) and Monte Carlo (MC) simulations of atomistic models (effective Hamiltonians~\cite{zhong1995}, all-atom force fields~\cite{shin2005development}, etc.) can probe larger scales but rely on empirically chosen ansatze, making accuracy difficult to assess and improve systematically. Consequently, most studies assume {\it a priori} CS~\cite{takenaka2017slush, akbarzadeh2012finite} for complex perovskites or optimize CS with semi-quantitative models~\cite{zhang2021compositional}.
Experimental probes of CS are also limited~\cite{kopecky2016nanometer, cabral2018gradient, eremenko2019local}. For example, X-ray scattering reconstructs the averaged electron-density distribution, providing only statistical correlations of ionic distributions~\cite{kopecky2016nanometer}.
Annular dark-field scanning transmission electron microscopy (STEM)~\cite{pennycook1988chemically} can distinguish cations in real space, but projection effects and sample thickness obscure three-dimensional (3D) CS~\cite{cabral2018gradient}.
Reverse Monte Carlo fits to X-ray and neutron scattering data may reproduce average geometric statistics and infer 3D CS~\cite{eremenko2019local}, but the accuracy of refined CS is questionable because the mapping from scattering measurements to chemical order is subtle and non-unique~\cite{SupplementalMaterial}.

To address this gap, we develop a first-principles framework that couples FIRE-Swap, a hybrid structure-sampling algorithm, with machine-learning interatomic potentials (MLIPs) to systematically determine intrinsic CS in complex perovskites. The framework is model-agnostic rather than tied to a specific MLIP architecture. In the following, we first introduce the algorithm and then demonstrate it using representative lead-based complex perovskites, including relaxor and non-relaxor solid solutions.

{\it Methods.} Our approach requires MLIPs that represent the first-principles potential energy surface (PES) $U(\mathbf{s}, \mathbf{R})$, where $\mathbf{s}=(s_1, \cdots, s_N)$ and $\mathbf{R}=(\mathbf{r}_1, \cdots, \mathbf{r}_N)$ denote the species and positions of all $N$ atoms in a periodic perovskite supercell.
The MLIPs are trained on extensive DFT data that covers local compositional and geometric disorder, either for a single complex perovskite (dedicated MLIP) or across multiple complex perovskites (universal MLIP).
Inspired by hybrid MD/MC methods for accelerated sampling~\cite{neyts2013combining, widom2014hybrid, antillon2020chemical}, FIRE-Swap (see Fig.~\ref{fig:workflow}) uses an MLIP to alternate between (a) geometric optimization with the Fast Inertial Relaxation Engine (FIRE) algorithm~\cite{bitzek2006structural} and (b) compositional relaxation via MC swapping of the chemical species of a random pair of target atoms, such as two A-site (or B-site) atoms in a complex perovskite with mixed occupancy.

Specifically, for the $n$-th iteration, step (a) seeks the inherent structure $\mathbf{R}'_n = \text{argmin}_{\mathbf{R}} U(\mathbf{s}'_n, \mathbf{R})$ for a proposed new CS $\mathbf{s}'_n$ that differs from the previous CS $\mathbf{s}^*_{n-1}$. Step (b) computes the swap energy cost: $\Delta E_n = U(\mathbf{s}'_n, \mathbf{R}'_n) - U(\mathbf{s}^*_{n-1}, \mathbf{R}^*_{n-1})$. When $\Delta E_n < 0$, we accept the swap by setting $\mathbf{R}^*_n = \mathbf{R}'_n$ and $\mathbf{s}^*_n = \mathbf{s}'_n$. When $\Delta E_n > 0$, we accept the swap with probability $p = e^{-\Delta E_n/k_BT_{\mathrm{FS}}}$, where $T_{\mathrm{FS}}$ is the sampling temperature. If rejected, we keep $\mathbf{R}^*_n = \mathbf{R}^*_{n-1}$ and $\mathbf{s}^*_n = \mathbf{s}^*_{n-1}$. We then propose a new configuration $\mathbf{s}'_{n+1}$ by swapping an arbitrary pair of target atoms that are nearest or next-nearest neighbors in $\mathbf{s}^*_n$. Neighborhood is defined by graph distance on the simple-cubic lattice. Each A site has 6 nearest and 12 next-nearest A-site neighbors. The same rule applies to B sites.
The approach generalizes straightforwardly to systems with both mixed A-site and B-site occupancy.

Unlike hybrid MD/MC methods~\cite{neyts2013combining}, which sample from a continuous distribution of atomic configurations, FIRE-Swap samples the equilibrium distribution $p_{\text{in}}$ over the discrete set of inherent structures (local minima of the PES; see Fig.~\ref{fig:workflow}(b))~\cite{stillinger1983inherent, corti1997constraints, nakagawa2006inherent}. It is important to choose $T_{\mathrm{FS}}$ sufficiently below the melting temperature $T_{\text{M}}$ so that $p_{\text{in}}$ is concentrated on low-energy inherent structures, i.e., intrinsic structures. But $T_{\mathrm{FS}}$ should not be too low, or the simulation may fail to escape mid-lying energy minima within a reasonable time, especially because the initial structure is typically high in energy. In practice, if the goal is to instead elucidate the CS of materials synthesized at high temperature and quenched without full equilibration, one may choose $T_{\mathrm{FS}}$ close to $T_{\text{M}}$ to enhance sampling of mid-lying CS favored by entropy.

{\it Results.}
We apply FIRE-Swap to three lead-based complex perovskites: the relaxor Pb$(\text{Mg}_{1/3}\text{Nb}_{2/3})\text{O}_3$ (PMN) and the ferroelectric solid solutions Pb$(\text{Zr}_{1/3}\text{Ti}_{2/3})\text{O}_3$ (PZT) and $(\text{Pb}_{1/3}\text{Sr}_{2/3})\text{TiO}_3$ (PST).
We use two density functionals, PBEsol~\cite{perdew2008restoring} and SCAN~\cite{sun2015strongly}, because they are among the best-performing GGA- and meta-GGA-level choices, respectively, for ferroelectric perovskites~\cite{scan2017}. We then employ four MLIPs: a SCAN-based Deep Potential (DP) model~\cite{han2017deep, zhang2018deep, zhang2018end}, a SCAN-based Cartesian Atomic Cluster Expansion Long-Range (CACE-LR) model~\cite{cheng2024cartesian, king2025machine, zhong2025machine}, a PBEsol-based CACE-LR model, and a PBEsol-based UniPero model~\cite{wu2023universal}.
The DP and CACE-LR models are dedicated MLIPs that we trained for PMN, whereas UniPero is an existing universal MLIP for complex perovskites involving 14 metal elements and their solid solutions, including PMN, PZT, and PST~\cite{wu2023universal}.
The CACE-LR models include long-range electrostatic interactions, with Latent Ewald Summation~\cite{cheng2025latent}. The DP model is numerically more efficient than the others. All models achieve energy prediction errors on the order of 1 meV per atom on their respective test sets. Their training sets consist of small complex-perovskite supercells with different compositional and geometric disorder. Details are provided in Appendix A.

\begin{figure}[bt]
  \centering
  \includegraphics[width=\linewidth]{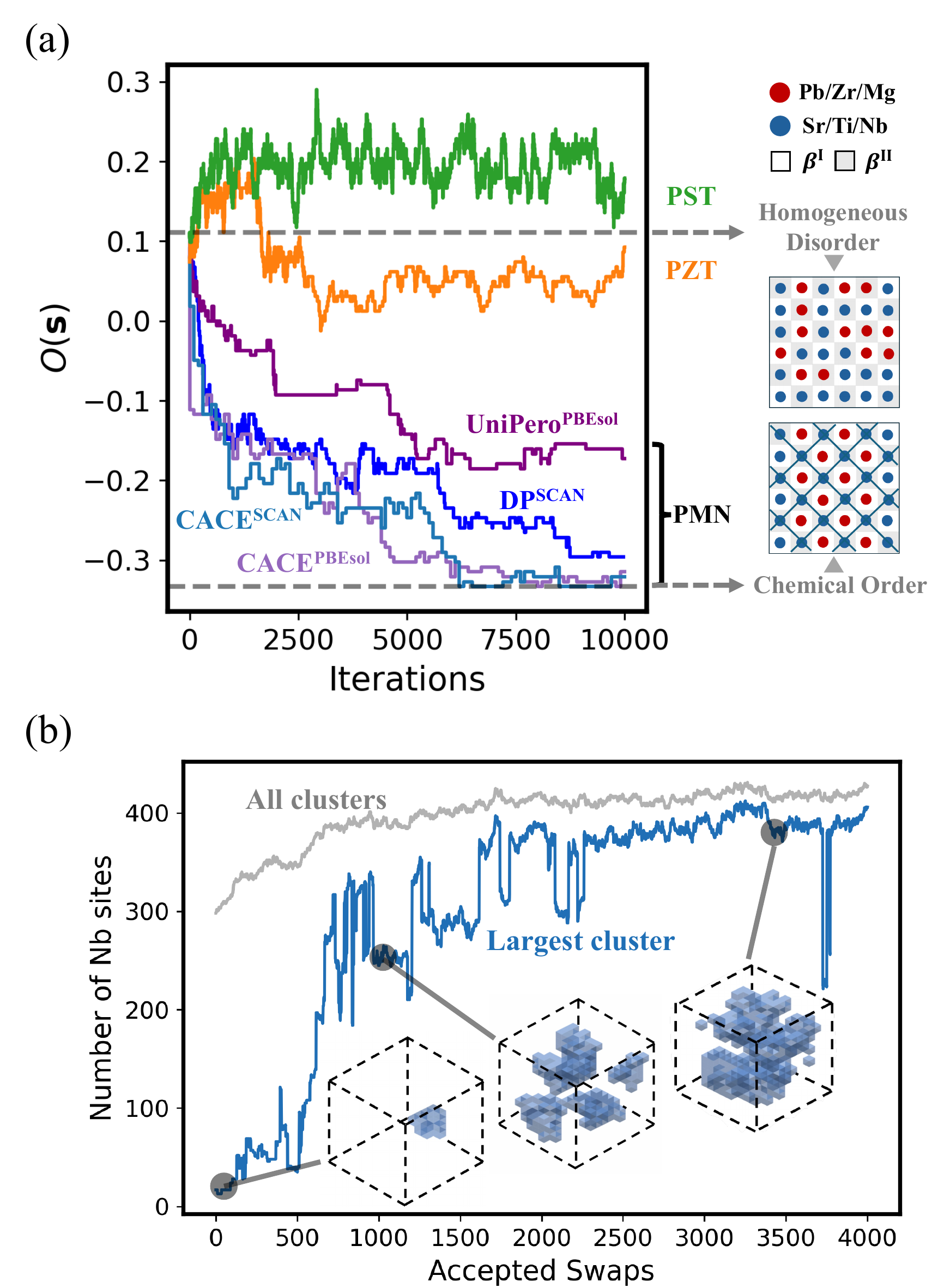}
  \caption{(a) Order parameter $O(\mathbf{s})$ as a function of FIRE-Swap iterations on the $6 \times 6 \times 6$ supercell. The right panel illustrates cross sections of the simulation cell, corresponding to CS with homogeneous disorder (PST and PZT) and CS with a chemical order (PMN). Sublattice $\beta^{\mathrm{II}}$ is represented by gray tiles.  (b) The number of Nb sites in all Nb clusters (gray) and in the largest cluster (blue). The insets show the geometry of the largest cluster in the $12 \times 12 \times 12$ lattice. }
  \label{fig:optimization}
\end{figure}

It is broadly accepted that PZT and PST lack significant chemical order, whereas the extent of chemical order in PMN remains debated~\cite{chen1989ordering, randall1990classification, viehland1991glassy, boulesteix1994numerical, kutnjak1999slow, davies2000chemical, akbas2000thermally, xu2000direct, farber2003influence, fu2009relaxor, cabral2018gradient}.
Two models have dominated discussions of PMN CS. The first is the charge-balanced random-site model~\cite{davies2000chemical} (also known as the random-layer model), which involves two interpenetrating face-centered-cubic sublattices (denoted $\beta^{\mathrm{I}}$ and $\beta^{\mathrm{II}}$; see Fig.~\ref{fig:optimization}(a)) in the perovskite structure. In this model, B sites in $\beta^{\mathrm{II}}$ are occupied almost exclusively by Nb ions~\footnote{The formation of Nb-rich sublattice prevents the adjacency of the Mg sites that presumably contribute to greater electrostatic repulsion than the adjacent Mg-Nb and Nb-Nb sites. }, whereas $\beta^{\mathrm{I}}$ contains a random 1:2 mixture of Nb and Mg ions. The second is the space-charge model~\cite{chen1989ordering}, in which $\mathrm{PbNb_{1/2}Mg_{1/2} O_3}$ rock-salt-ordered clusters (for brevity, rock-salt clusters) coexist with $\mathrm{PbNb O_3}$ clusters (Nb clusters). Because both clusters carry net space charge, rock-salt clusters were postulated to be nanodomains with Nb-ion shells. Over the years, experiments have supported the existence of the Nb-rich $\beta^{\mathrm{II}}$ structure~\cite{chen1989ordering, cantoni2004direct, cabral2018gradient, kumar2021atomic}, but concrete evidence for the space-charge model remains lacking. Electrostatic considerations also disfavor the space-charge model. However, the random-site model is also incomplete: its assumption of a fully random $\beta^{\mathrm{I}}$ sublattice has not yet been confirmed and warrants re-examination.

Together, the four MLIPs enable a first-principles examination of the aforementioned issues and provide a strong consistency check. We first use UniPero for a comparative study of PMN, PZT, and PST, confirming that only PMN exhibits notable chemical order. We perform FIRE-Swap simulations on $6 \times 6 \times 6$ supercells of PMN, PZT, and PST at $T_{\mathrm{FS}}=600$ K, 300 K, and 300 K, respectively (see Appendix B for technical details). Each simulation starts from a homogeneously disordered configuration, with mixed B-site atoms (PMN and PZT) or mixed A-site atoms (PST) randomly assigned to the lattice. To track formation of a sublattice dominated by one element, we use the order parameter $O(\mathbf{s})=\frac{1}{6N_c} \sum_{i\sim j} s_i s_j$, where $N_c=216$ is the total number of primitive sites and $i\sim j$ denotes all nearest-neighbor pairs of B sites (PMN and PZT) or A sites (PST). We set $s_i=1$ for Sr, Ti, and Nb, and $s_i=-1$ otherwise. A homogeneously disordered state gives $O=1/9$, whereas $O=-\tfrac{1}{3}$ corresponds to a $\beta^{\mathrm{II}}$ sublattice fully occupied by one element. All simulations equilibrate within $10^4$ iterations.

Results are shown in Fig.~\ref{fig:optimization}(a). Indeed, $O$ fluctuates around $1/9$ for PZT and PST, consistent with homogeneous disorder, but approaches $-\tfrac{1}{3}$ for PMN. All models except UniPero predict a stable $O$ near $-\tfrac{1}{3}$, with cumulative $\Delta E_n$ from accepted swaps of 60--70 meV/f.u. in each case. UniPero predicts a higher equilibrium $O$ because small Mg clusters ($\mathrm{PbMgO_3}$ units) locally disrupt chemical order in Nb-rich $\beta^{\mathrm{II}}$. The discrepancy between PBEsol-based CACE-LR and UniPero likely reflects missing long-range electrostatic interactions in UniPero, which effectively captures only short-range electrostatics within $6~\text{\AA}$~\cite{wu2023universal}. This bias may facilitate Mg clustering, which should otherwise be disfavored because the average ionic charge of a $\mathrm{PbMgO_3}$ unit is $-2e$, whereas that of a $\mathrm{PbNbO_3}$ unit is only $+e$.

A comprehensive study of long-range electrostatics is left for future work. Within the present scope, the key result is robust: across different density functionals and MLIP architectures, FIRE-Swap reveals the Nb-rich sublattice prescribed by the random-site model, distinguishing the intrinsic CS of PMN from those of PZT and PST solid solutions. Furthermore, we perform DP-driven FIRE-Swap simulations for PMN at $T_{\mathrm{FS}}=1200$ K, comparable to experimental synthesis temperatures. We find that $O$ fluctuates between $-0.15$ and $-1/3$, rather than steadily approaching $-1/3$~\cite{SupplementalMaterial}. This deviation from $-1/3$ arises from strong thermal fluctuations that interrupt long-range chemical order, producing small chemically disordered regions. These features are consistent with STEM evidence of interrupted chemical order and anti-phase boundaries~\footnote{Anti-phase boundaries are disordered regions where the $\beta^{\text{I/II}}$ classification flips.} in PMN samples~\cite{cabral2018gradient}. 

\begin{figure}[bt]
  \centering
  \includegraphics[width=0.9\linewidth]{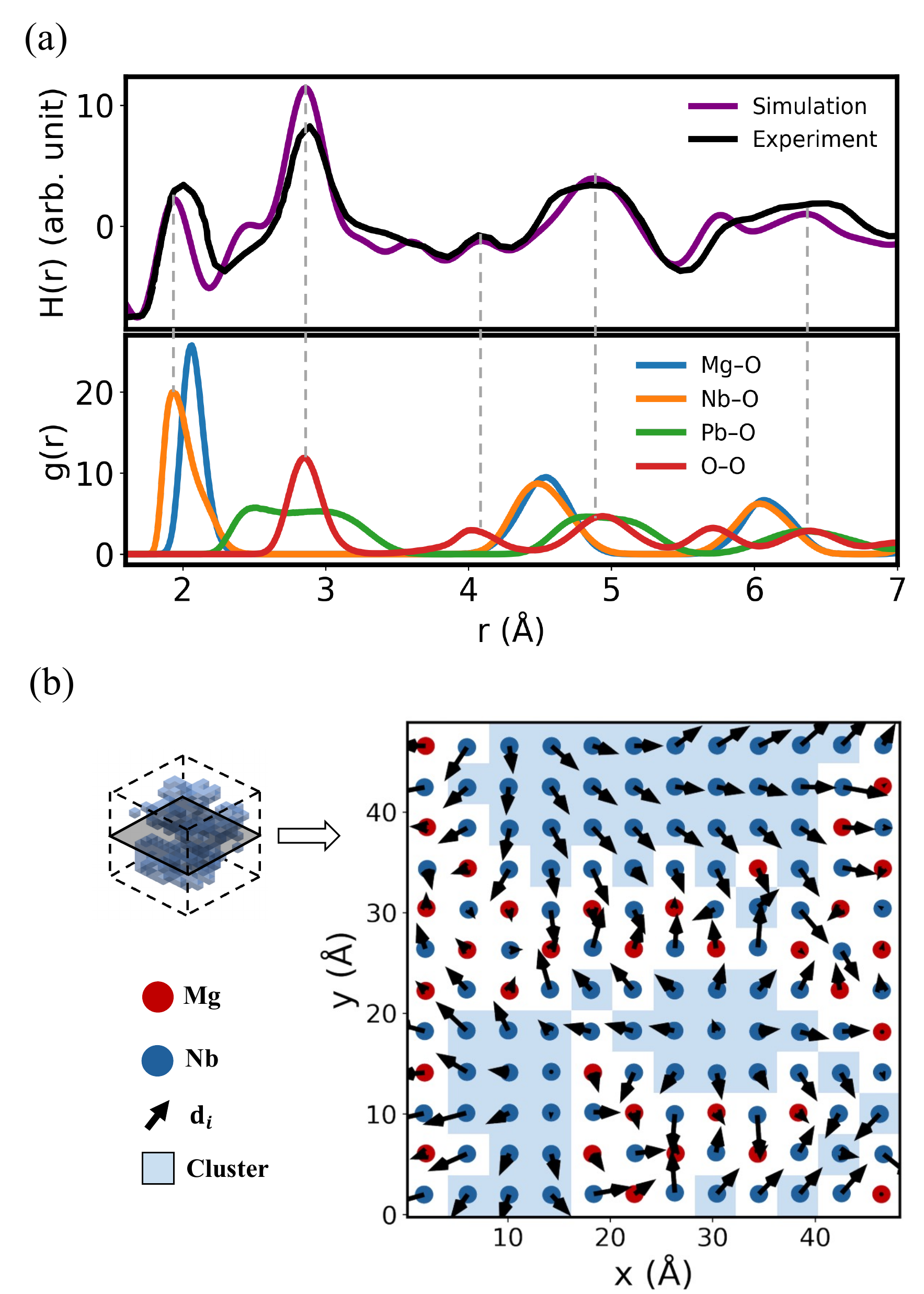}
  \caption{(a) Upper panel: The predicted (purple) and experimental~\cite{eremenko2019local} (black) $H(r)$ associated with PMN powder samples at $T=300$K. Lower panel: The predicted radial distribution function for $T=300$K.   (b) ($d_i^x, d_i^y$) as arrows on an arbitrary perovskite layer (in the XY plane) of the simulation box. Blue shades mark Nb clusters. }
  \label{fig:RDF}
\end{figure}
Next, FIRE-Swap simulations of a $12 \times 12 \times 12$ supercell ($N_c=1728$) reveal larger-scale features of PMN CS, including nontrivial Nb clustering that goes beyond the random-site model. We define a Nb cluster as follows: a Nb site belongs to a cluster if it has at most one Mg-site neighbor among its six nearest neighbors. All such Nb sites form a lattice subgraph, which can be uniquely partitioned into connected components (clusters). Different clusters are disconnected on the lattice. The cluster size is the number of Nb sites in each component. For any given CS, a Breadth-First Search algorithm~\cite{moore1959shortest, lee2009algorithm} computes the Nb-cluster size distribution and identifies the largest cluster. To accelerate sampling, we initialize CS according to the random-site model. The initial CS contains about 300 Nb sites distributed across roughly 250 Nb clusters.
Figure~\ref{fig:optimization}(b) shows results from the DP-driven simulation at $T_{\mathrm{FS}}=600$ K, in which about 4000 swaps are accepted over $3\times 10^5$ iterations. The cumulative $\Delta E_n$ from accepted swaps is about 20 meV/f.u.
The total number of Nb sites in all Nb clusters increases to about 430 (gray curve), close to the upper bound $N_c/3$~\footnote{The upper bound corresponds to maximal separation between Nb clusters and rock-salt clusters. The upper bound cannot be reached because our definition of a Nb cluster omits sites at the interface.}. The largest Nb cluster grows from 17 to about 400 sites (blue curve). We find almost all clustered Nb sites collapse into a single cluster that almost percolates the lattice. Nb sites outside Nb clusters are mostly embedded in the rock-salt cluster~\footnote{We verify this by tracking the number of Nb sites in the rock-salt cluster, where an Nb site belongs to the rock-salt cluster if it has at most one Nb neighbor. Details are omitted here but available in the Supplementary Data and Code~\cite{pmn-git}.} and belong to sublattice $\beta^{\mathrm{II}}$. The chemical ordering in $\beta^{\mathrm{II}}$ is well maintained: $O(\mathbf{s})$ remains within $[-1/3,-0.3]$. Throughout the simulation, one third of nearest-neighbor pairs are Nb--Nb, two thirds are Nb--Mg, and less than $1\%$ are Mg--Mg. These values agree well with X-ray-scattering statistics: $33.7\%$ of nearest-neighbor pairs are Nb--Nb and $65.8\%$ are Nb--Mg~\cite{kopecky2016nanometer}.

We identify two anomalies in Nb clustering relative to standard phase-separation behavior. First, percolation theory on $\beta^{\mathrm{I}}$ predicts a power-law cluster-size distribution, which would naively imply many more intermediate-size clusters. Second, the surface of the largest Nb cluster does not coarsen substantially (see illustrations in Fig.~\ref{fig:optimization}(b)). Its surface-to-volume ratio $\eta$ remains above 2 throughout the simulation, whereas a perfect cubic cluster with volume $N_c/3$ would give $\eta<1$.
The first anomaly is expected: Nb ions anchored in $\beta^{\mathrm{II}}$ strongly promote the formation of a dominant Nb cluster, rendering standard percolation arguments irrelevant.
The second anomaly may arise because local interactions favor 1:1 rock-salt ordering and maximize the Nb--Mg interfacial area. As a result, the largest Nb cluster adopts a mesh-like geometry with high $\eta$, which suppresses space-charge accumulation and eliminates the need for the Nb-ion shell hypothesized in the space-charge model.
These observations go beyond existing CS models and provide a first-principles mesoscale description of PMN, which we term the ``anchored-mesh'' model.

UniPero- and CACE-LR-driven FIRE-Swap simulations of the same PMN supercell consistently support the formation of a dominant Nb cluster~\cite{SupplementalMaterial} and the anchored-mesh model. Finite-size artifacts are also ruled out~\cite{SupplementalMaterial}. By contrast, FIRE-Swap simulations of a $12 \times 12 \times 12$ PZT supercell, the isovalent counterpart of PMN, show a lack of chemical order and numerous small Zr and Ti clusters~\cite{SupplementalMaterial}.

How do the anchored-mesh model and the MLIP together describes geometric and dipolar properties of PMN? We perform DP-driven NVT MD simulations of a $12 \times 12 \times 12$ PMN supercell with the optimized intrinsic CS over thermal temperatures $T$ from 100 K to 700 K, with each run lasting 10 ns.
For $T=300$ K, we calculate the spherically symmetrized structure factor $S(Q)$ using standard neutron scattering lengths~\footnote{Pb: 9.405 fm, Mg: -3.73 fm, Nb: 7.054 fm, O: 5.803 fm}. The pair distribution function $H(r)\propto r^{-1}\int_0^\infty Q(S(Q)-1)\sin(Qr)dQ$ is plotted in Fig.~\ref{fig:RDF}(a) for comparison with measurements on PMN powder samples~\footnote{See Supplementary Figure 1 of Ref.~\cite{eremenko2019local}.}. The overall agreement is decent. A minor but notable discrepancy appears near $r\approx 2.5~\text{\AA}$, where the predicted $H(r)$ shows a small peak that is absent in experiment. The origin of peaks in $H(r)$ is illustrated in the lower panel of Fig.~\ref{fig:RDF}(a), which reports the radial distribution function $g(r)$. We find that the small peak near $r\approx 2.5~\text{\AA}$ in the predicted $H(r)$ is associated with Pb-O pairs, which exhibit a broad distribution over $[2.25,3.5]~\text{\AA}$. This is in close agreement with the broad Pb-O peaks (one at $r\approx 2.4~\text{\AA}$ and strengthened below room temperature) in the same interval from X-ray scattering~\footnote{See Supplementary Figure 17(a) of Ref.~\cite{eremenko2019local}.}. These minor discrepancies are primarily due to errors in density functional approximation~\cite{scan2017}, which affect bond lengths and thermal broadening. In ~\cite{SupplementalMaterial}, we show that both $H(r)$ and $g(r)$ are not sensitive to changes in CS~\footnote{This highlights potential non-uniqueness in CS inferred from reverse Monte Carlo methods.}.

Then, we study how CS shapes polar nanoregions (PNRs) in relaxors~\cite{wakimoto2002ferroelectric, blinc2003field, jeong2005direct, gehring2009reassessment, fu2009relaxor, kumar2021atomic}. We characterize local polarity in PMN using the displacement $\mathbf{d}_i$ of each B-site ion $i$ from the geometric center of its oxygen octahedral cage. $\mathbf{d}_i$ is strongly correlated with local (primitive-cell) dipole moments~\footnote{The local dipole moment can be rigorously defined from the Wannier-center representation of polarization prescribed by Berry-phase theory~\cite{Resta2007, RevModPhys.84.1419}, as in Ref.~\cite{pto2025prb}.}, and the softening of its vibration is a signature of ferroelectric phase transitions~\cite{rabe2007physics}. The displacement of the A-site Pb ion relative to neighboring oxygen ions can also describe local polarity~\cite{takenaka2017slush, eremenko2019local}, but it is less informative for understanding the role of chemical heterogeneity~\cite{SupplementalMaterial}. A temperature-dependent characterization of the probability density function (PDF) of $\mathbf{d}_i$ is reported in Appendix D, showing how dipolar disorder quenches as temperature decreases from 700 K to 100 K, in agreement with the phenomenological picture of different stages of relaxor ferroelectricity~\cite{bokov2020recent}.

Here, we focus on the direct connection between CS and PNRs. In Fig.~\ref{fig:RDF}(b), we report a representative configuration of $\mathbf{d}_i$ using a snapshot of $(d_i^x, d_i^y)$ on one layer (in the XY plane) of the supercell from the $T=100$ K MD simulation. At low temperatures, the direction of $\mathbf{d}_i$ is nearly frozen against thermal fluctuations, forming clear PNRs within the Nb clusters marked with blue shading in Fig.~\ref{fig:RDF}(b). Neighboring $\mathbf{d}_i$ vectors in the Nb clusters appear strongly correlated but not rigidly aligned, and are therefore potentially subject to reorientation under thermal and electric perturbation. We find that the phenomenological picture of individual spherical nanosized PNRs~\cite{fu2009relaxor, fu2013pb} is likely imprecise.
In the anchored-mesh model, PNRs composed of correlated Nb sites are not separated by a disordered matrix and do not behave as discrete, frustrated ``mesoscopic dipoles.'' Instead, they form interconnected, non-coarsened dipolar structures that should be studied as a collective whole. This motivates future investigation of mesoscale mechanisms that drive fluctuations of PNRs.

In closing, our approach has been applied to multiple complex perovskites and reveals sharp compositional differences between PMN and isovalent solid solutions. The anchored-mesh model provides new insights into the intrinsic CS of PMN, reveals further complexity in PNR morphology and calls for reexamination of the mechanism underlies the unique dielectric responses of relaxors. These studies are carried out on first-principles grounds without uncontrolled phenomenological assumptions. With recent progress in constructing accurate universal MLIPs~\cite{deng2023chgnet, zhang2024dpa, batatia2025foundation}, our approach can be readily applied to large-scale studies of existing complex perovskites and those yet to be discovered. To facilitate this, we developed the open-source Python package \texttt{PeroStruc}~\cite{perostruc}, which enables FIRE-Swap simulations with any MLIP that interfaces with the calculator class in the \texttt{Atomic Simulation Environment}~\cite{hjorth2017atomic} library.

{\bf Data and Code Availability} --
Data sets, models, and scripts that can reproduce the findings of this study are publicly available on GitHub~\cite{pmn-git}.

{\bf Acknowledgement} --
We thank Roberto Car for fruitful discussions. VASP calculations were done when P.X. and K.C. were graduate students in Roberto Car's group at Princeton, where the VASP license is issued. P.X. and K.C. were supported by the Computational Chemical Sciences Center: Chemistry in Solution and at Interfaces (CSI), funded by the US Department of Energy (DOE) Award DE-SC0019394. Part of calculations were performed using the Princeton Research Computing resources at Princeton University, which is a consortium of groups led by the Princeton Institute for Computational Science and Engineering (PICSciE) and Office of Information Technology's Research Computing. P.X. was then supported by the Alvarez Fellowship of Lawrence Berkeley National Lab. X.X. was supported by the Berkeley Lab 2025 Summer Program. X.X. and P.X. were supported by the DOE Advanced Scientific Computing Research (ASCR) Applied Mathematics program under Contract No. DE-AC02-05CH11231. P.Z. was supported by the NUS Presidential Young Professorship and the computational resources from NUS-HPC (CFP04-CF-010).
This research used resources of the National Energy Research Scientific Computing Center (NERSC), a Department of Energy User Facility, using AI4Sci@NERSC award DDR-ERCAP 34636.

\bibliography{references}

\section*{Appendices}
\textbf{Appendix A - Details of MLIPs}

{\it Deep Potential.} Our Deep Potential (DP) model for PMN is based on DFT with SCAN approximation~\cite{sun2015strongly}. The combination of DP and SCAN has also been used to model other ferroelectrics ~\cite{yang2024deuteration, pto2025prb, xie2026ab}. We use \texttt{VASP}~\cite{kresse1996iter,kresse1996efficiency} and the projector augmented wave potentials \cite{kresse1999ultrasoft} for all DFT calculations.
The smallest spacing between k-points is 0.6 $\text{\AA}^{-1}$. Gamma point is always included.
The SCAN-DFT dataset is generated using the active-learning framework \texttt{DPGEN}~\cite{zhang2019active, zhang2020dp}, with pressures in the interval $[1, 3\times 10^4]$ bar and temperatures in the interval $[150, 650]$ K. Active exploration of the configuration space is performed via NPT-MD simulations initialized from PMN configurations with different sizes and disorder patterns. Specifically, three types of initial configurations are used: (1) charge-balanced supercells of 36 PMN units (180 atoms) prescribed by the random-site model; (2) charge-balanced supercells of 72 PMN units prescribed by the random-site model; and (3) charge-balanced supercells of 27 PMN units with complete B-site disorder.

The active learning procedure consists of 15 iterations of explorations. We used data from 12 iterations, together with the initial dataset manually generated before active learning, as the train set, which contains DFT energy and force labels for 1386 different atomic configurations. The data from the other 3 iterations are used as the test set, containing 300 different atomic configurations. With the train set, we use \texttt{DeePMD-Kit}~\cite{wang2018deepmd, LU2021107624, zeng2023deepmd} to train a DP model with a short-range cutoff $6$\AA.
A plot of model error distribution in the train set and the test set is provided in ~\cite{SupplementalMaterial}. We find a root-mean-squared error (RMSE) of 0.8 meV/atom and 1.0 meV/atom on the train set and the test set, respectively.

{\it CACE-LR.} The CACE-LR model captures long-range electrostatic interactions through latent Ewald summation~\cite{cheng2025latent, zhong2025machine, king2025machine}. In this work, the CACE-LR model is trained for Pb(Mg$_{1/3}$Nb$_{2/3}$)$_{1-x}$Ti$_x$O$_3$ solid solutions following the Pretrain-FineTune-Distillation (\texttt{PFD-kit}) \cite{wang_pre-training_2025} protocol. Here, the pretrained baseline model is selected as the universal DPA-3 model~\cite{zhang2025graph}. We use 2,000 PBEsol-DFT data points to fine-tune the DPA-3 model, covering six composition ratios ($x=0, 0.25, 0.33, 0.4, 0.5, 1.0$) and various Mg/Nb/Ti compositional structure. These data points are selected from NPT-MD (700K) 
trajectories using information-entropy analysis. The fine-tuned, short-ranged DPA-3 model reaches RMSEs of 0.49 meV/atom for energy and 27 meV/\AA{} for force. We then perform NPT-MD simulations with this DPA-3 model. We select and label 10,000 configurations from MD trajectories to construct the final PBEsol-DFT dataset for CACE-LR training. The final CACE-LR model achieves accuracies of 1.15 meV/atom for energy and 40 meV/\AA{} for force relative to DFT data.

{\it UniPero.}
UniPero is built on the DPA-1 (Deep Potential with self-attention)~\cite{zhang2024pretraining} framework and trained on first-principles data of perovskites involving 14 metal elements and their solid solutions over arbitrary compositions. UniPero has a short-range cutoff $6$\AA. A complete description of the training dataset, data-generation workflow, DFT labeling protocol, and training erros is provided in Ref.~\cite{wu2023universal}, to which we refer readers for full details.

\textbf{Appendix B - Details of FIRE-Swap simulation}


\textit{PMN.} The DP-driven simulations are performed with fixed lattice constant $a=4.05~\text{\AA}$. In each FIRE relaxation, the force convergence threshold and the maximum number of optimization steps are set to $f_{\mathrm{thr}}=0.02~\mathrm{eV}/\text{\AA}$ and $N_{\mathrm{step}}=100$, respectively.

The UniPero-driven simulations are performed with fixed lattice constant $a=4.05~\text{\AA}$. We use $T_{\mathrm{FS}}=600$ K, $f_{\mathrm{thr}}=0.03~\mathrm{eV}/\text{\AA}$ and $N_{\mathrm{step}}=100$ for $6\times6\times6$ supercell. We use $T_{\mathrm{FS}}=450$ K, $f_{\mathrm{thr}}=0.03~\mathrm{eV}/\text{\AA}$ and $N_{\mathrm{step}}=50$ for $12\times 12\times 12$ supercell.

The CACE-LR-driven simulations are performed with flexible cell parameters. We use $T_{\mathrm{FS}}=600$ K, $f_{\mathrm{thr}}=0.02~\mathrm{eV}/\text{\AA}$ and $N_{\mathrm{step}}=200$.

\textit{PZT.} Simulations are performed with flexible cell parameters. We use $T_{\mathrm{FS}}=300~\mathrm{K}$, $f_{\mathrm{thr}}=0.02~\mathrm{eV}/\text{\AA}$ and $N_{\mathrm{step}}=100$.

\textit{PST.} Simulations are performed with flexible cell parameters and $T_{\mathrm{FS}}=300~\mathrm{K}$.  We use $f_{\mathrm{thr}}=0.03~\mathrm{eV}/\text{\AA}$ and $N_{\mathrm{step}}=100$ for $6\times 6\times 6$ supercell. We use $f_{\mathrm{thr}}=0.03~\mathrm{eV}/\text{\AA}$ and $N_{\mathrm{step}}=150$ for $12\times 12\times 12$ supercell.

\textbf{Appendix C - Energy scales in FIRE-Swap simulations}

\begin{figure}[bt]
  \centering
  \includegraphics[width=\linewidth]{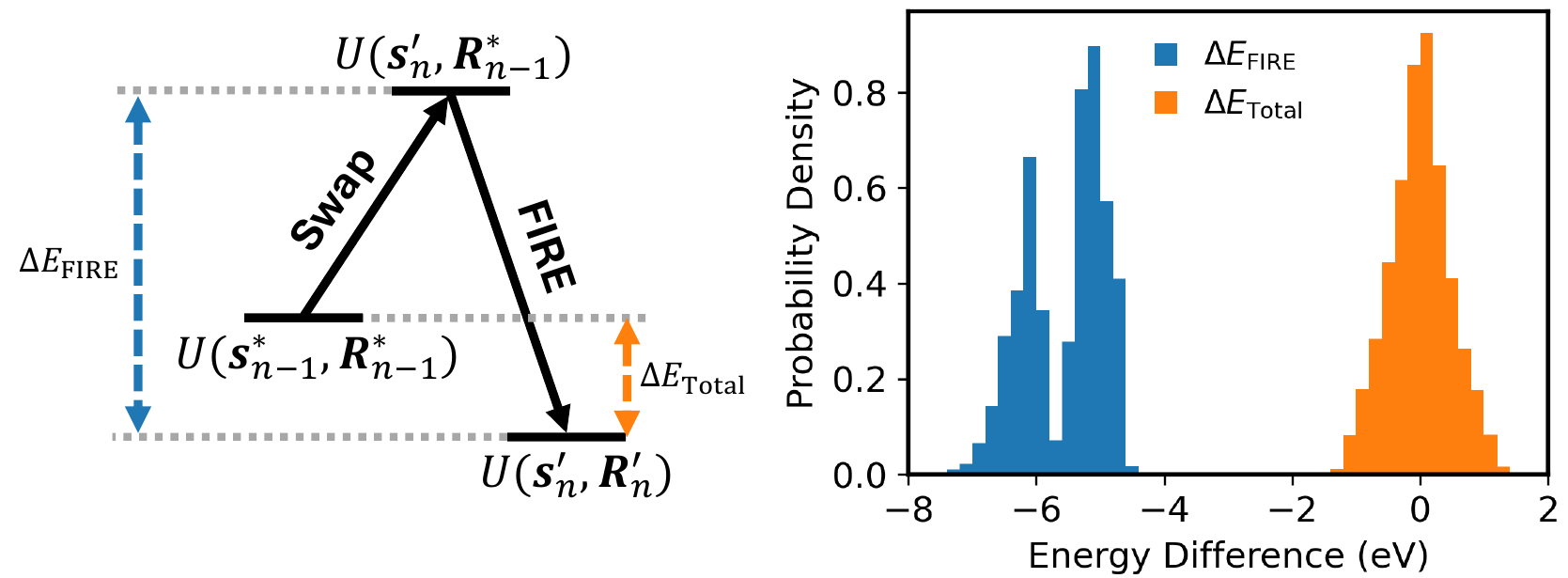}
  \caption{Left panel: A sketch of energy changes, $\Delta E_{\text{FIRE}}$ and $\Delta E_{\text{Total}}$, in one iteration of FIRE-Swap. Right panel: the probability density distribution of $\Delta E_{\text{FIRE}}$ and $\Delta E_{\text{Total}}$ in a FIRE-Swap simulation with around 10000 iterations. }
  \label{Fig:energy_dist}
\end{figure}

Here we show that the contribution of atomic geometric disorder to system energy is comparable to that from compositional disorder, which motivates the FIRE-Swap approach that treats compositional and geometric relaxations concurrently.
We focus on two quantities: $\Delta E_{\text{FIRE}}=U(\mathbf{s}'_n,  \mathbf{R}'_n )-U(\mathbf{s}'_n,  \mathbf{R}^*_{n-1})$ and
$\Delta E_{\text{Total}}=U(\mathbf{s}'_n,  \mathbf{R}'_n )-U(\mathbf{s}^*_{n-1},  \mathbf{R}^*_{n-1})$, as indicated in the left panel of Fig.~\ref{Fig:energy_dist}. The probability density distribution of $\Delta E_{\text{FIRE}}$ and $\Delta E_{\text{Total}}$ in the DP-driven FIRE-Swap simulation of a $6\times 6 \times 6$ PMN supercell is plotted in the right panel of Fig.~\ref{Fig:energy_dist}.

\begin{figure}[bt]
  \centering
  \includegraphics[width=\linewidth]{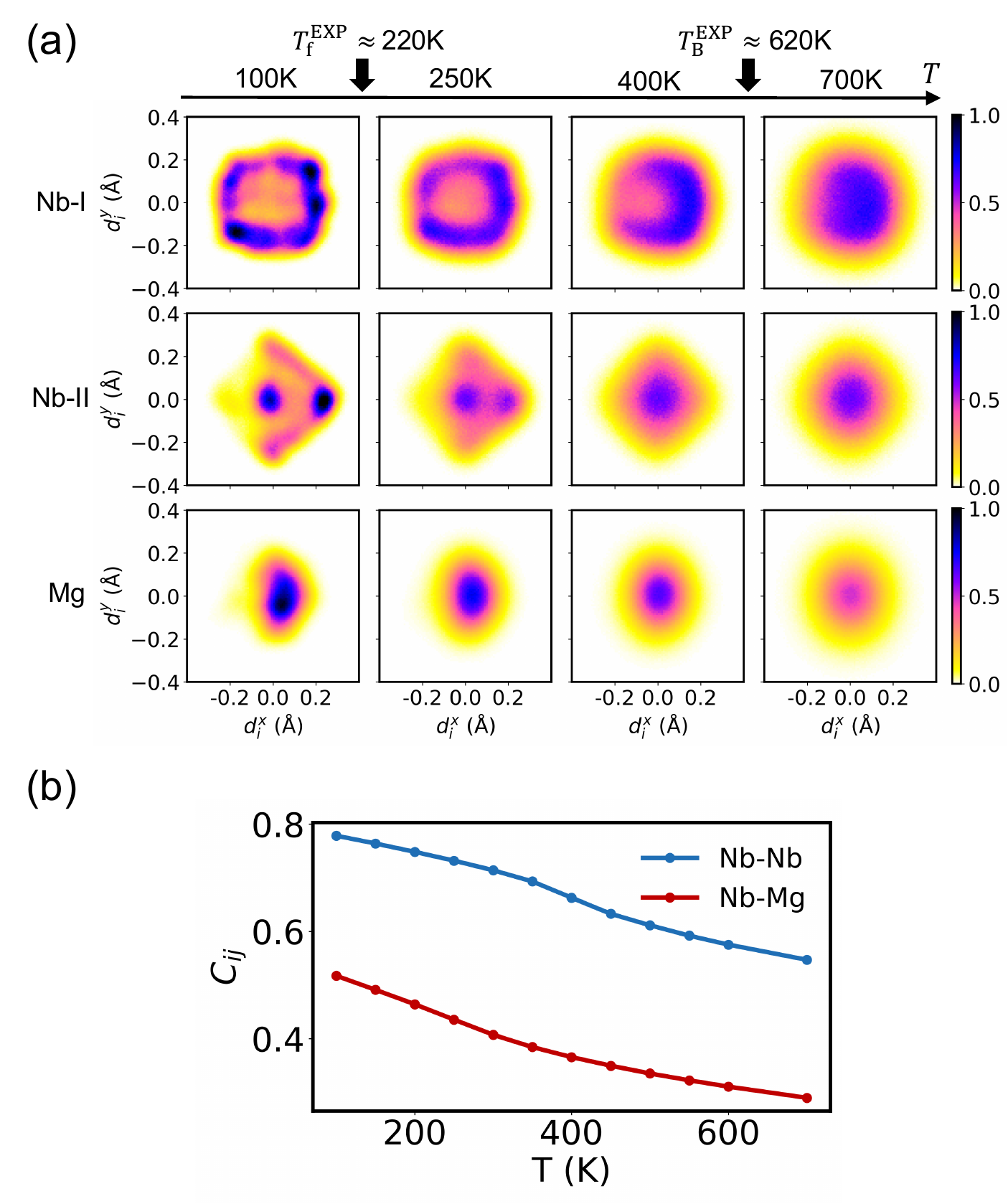}
  \caption{(a) Histograms of PDF $p(d_i^x, d_i^y)$ associated with different types of B sites and different temperatures. The PDF $p(d_i^x, d_i^z)$ has similar features and are omitted here. (b) The correlation functions $C_{ij}$ (averaged over eligible pairs) for nearest-neighbor Nb-Nb pairs and Nb-Mg pairs . }
  \label{fig:dipolar}
\end{figure}
The distribution of $\Delta E_{\text{FIRE}}$ is associated with the energy scale relevant to atomic geometric disorder. Let $\Delta E_{\text{Swap}}=\Delta E_{\text{Total}}-\Delta E_{\text{FIRE}}$. The distribution of $\Delta E_{\text{Swap}}$ is associated with the energy scale relevant to the compositional disorder.  From Fig.~\ref{Fig:energy_dist}, $\Delta E_{\text{FIRE}}$ is typically around -6 eV.  Meanwhile, the distribution of $\Delta E_{\text{Total}}$ resembles a zero-centered Gaussian with a standard deviation smaller than 1 eV. This suggests that the distribution of $\Delta E_{\text{Swap}}$ will be centered around 6 eV. In this case study, energy changes caused by swapping species and by geometric optimization are of the same order. It is the small remnant of their cancelation determines the acceptance of a proposed new CS.

\textbf{Appendix D - Dipolar properties of PMN}

We study temperature-dependent properties of $\mathbf{d}_i$ as an effective description of local polarity in PMN.
At low temperature, the behavior of $\mathbf{d}_i$ differs markedly across diffrent types of clusters. This sharp contrast is shown in Fig.~\ref{fig:dipolar}(a), which depicts the probability density function (PDF) $p(d_i^x, d_i^y)$ for Nb ions in Nb clusters (Nb-I), Nb ions in rock-salt clusters (Nb-II), and Mg ions.

Overall, these PDFs are consistent with the phenomenology of different stages of relaxor ferroelectricity. Above the Burns temperature, $T^{\text{EXP}}_{\text{B}}\approx 620~\mathrm{K}$~\cite{burns1983glassy}, the PDFs for Nb-I, Nb-II, and Mg all display spherical symmetry, indicating a paraelectric state. Between $T^{\text{EXP}}_{\text{B}}$ and the freezing temperature, $T^{\text{EXP}}_{\text{f}}\approx 220~\mathrm{K}$~\cite{viehland1991glassy}, signs of quenched disorder appear in $\mathbf{d}_i$ for Nb-I and Nb-II. Below $T^{\text{EXP}}_{\text{f}}$, this disorder in Nb-I and Nb-II becomes substantially quenched (i.e., frozen), corresponding to the so-called non-ergodic relaxor phase.
Meanwhile, the PDFs associated with Mg vary much less with temperature, indicating that polar fluctuations of Mg ions play a minor role in shaping dielectric properties.

Many qualitative differences between Nb-I and Nb-II can be identified from the PDFs. Some reflect finite-size effects caused by insufficient self-averaging, such as weak mirror-symmetry breaking (e.g., $x\rightarrow -x$) in the PDF, which is evident for $T\leq 250~\mathrm{K}$. This artifact can be reduced by simulating a much larger supercell.
Here, we focus on the key difference that is robust to finite-size effects and most relevant to how CS shapes relaxor ferroelectricity: at $T=100~\mathrm{K}$, the PDF of Nb-I exhibits a continuous distribution of disorder, whereas the PDF of Nb-II is less diffuse and shows separated peaks. This implies that Nb-II is less electrically susceptible than Nb-I because of its greater rigidity. The flexible $\mathbf{d}_i$ of Nb-I and its continuous distribution are reminiscent of PNRs, which are postulated to be a key mechanism of relaxor ferroelectricity.

In the main text, PNRs have been illustrated by Fig.~\ref{fig:RDF}(b). It is found that $\mathbf{d}_i$ in PNRs are strongly correlated but not rigidly alighed. Here, we quantify such correlation using the equilibrium correlation function for neighboring $\mathbf{d}_i$ and $\mathbf{d}_j$, defined as $C_{ij}=\langle \mathbf{d}_i \mathbf{d}_j \rangle/ \sqrt{\langle \|\mathbf{d}_i\|^2 \rangle \langle \|\mathbf{d}_j\|^2 \rangle}$. The results are shown in Fig.~\ref{fig:dipolar}(b). We found that the correlation function increases when temperature decreases. Notably, $C_{ij}$ for Nb pairs rises to almost 0.8 at $T=100$ K. This amounts to a substantial correlation length around 2nm. Meanwhile, $C_{ij}$ for Nb-Mg pairs that reside in rock-salt clusters has a correlation length small enough that different Nb-II sites are almost uncorrelated.
Presumably, dielectric responses of PMN depend on dynamical behaviors of PNRs. The strong but not rigid alighment of $\mathbf{d}_i$ potentially enables slow fluctuation of PNRs on the microsecond scale and above, leading to the unique Vogel-Fulcher law~\cite{viehland1990freezing} that describes frequency-dependent dielectric maximum in relaxor ferroelectrics. In principle, one can calculate the frequency-dependent dielectric permitivity $\epsilon(T)$ with our optimized CS. However, advanced enhanced sampling methods may be necessary to fill the huge gap between the typical time scale (ps$\sim$ns) in numerical simulation and that (ns$\sim$s) accessible to experiments. We leave this for future work.

\end{document}


\preprint{APS/123-QED}
\newcommand{\pto}{PbTi$\text{O}_3$ }
\newcommand{\bto}{BaTi$\text{O}_3$ }
\newcommand{\red}[1]{\textcolor{red}{#1}}
\newcommand{\blue}[1]{\textcolor{blue}{#1}}

\title{Supplemental Material for ``Intrinsic structure of relaxor ferroelectrics from first principles''}

\author{Xinyu Xu}
\affiliation{Applied Mathematics and Computational Research Division, Lawrence Berkeley National Laboratory, Berkeley, CA 94720, USA}
\affiliation{Department of Materials Science and Engineering, Stanford University, Stanford, CA 94305, USA}

\author{Kehan Cai}
\affiliation {Department of Chemistry, Princeton University, Princeton, NJ 08544, USA}

\author{Yubai Shi}
\affiliation{Department of Materials Science and Engineering, National University of Singapore, Singapore}

\author{Peichen Zhong}
\affiliation{Department of Materials Science and Engineering, National University of Singapore, Singapore}

\author{Pinchen Xie}
\affiliation{Applied Mathematics and Computational Research Division, Lawrence Berkeley National Laboratory, Berkeley, CA 94720, USA}
\date{\today}

\maketitle

\section{FIRE-Swap simulations of PMN close to the melting temperature}

Here we report simulation results of Deep Potential(DP)-driven FIRE-Swap for PMN at $T_{\mathrm{FS}}=1200$K, which is comparable to experimental synthesis temperatures.

The evolution of the order parameter $O(s)$ in the $6\times6\times6$ supercell of PMN is shown in Supplemental Fig.~\ref{fig:1200K}(a).  The equilibration is reached after roughly 2000 iterations. Then,  $O(s)$ fluctuates between  $-0.15$ and $-1/3$, rather than steadily approaching $-1/3$ as has been observed at lower $T_{\mathrm{FS}}$.
This fluctuating deviation from $-1/3$ leads to interruptions in long-range chemical order, as discussed in the main text.

The evolution of Nb cluster size in the $12\times12\times12$ supercell of PMN is shown in Supplemental Fig.~\ref{fig:1200K} (b). Again, the largest Nb cluster dominates the lattice after equilibration, and we find a corresponding $\eta$ around 3. However, due to strong thermal fluctuations, there are substantially more small Nb clusters compared to those in the simulation performed with $T_{\mathrm{FS}}=600$K.

\begin{figure}[h]
  \centering
  \includegraphics[width=0.8\textwidth]{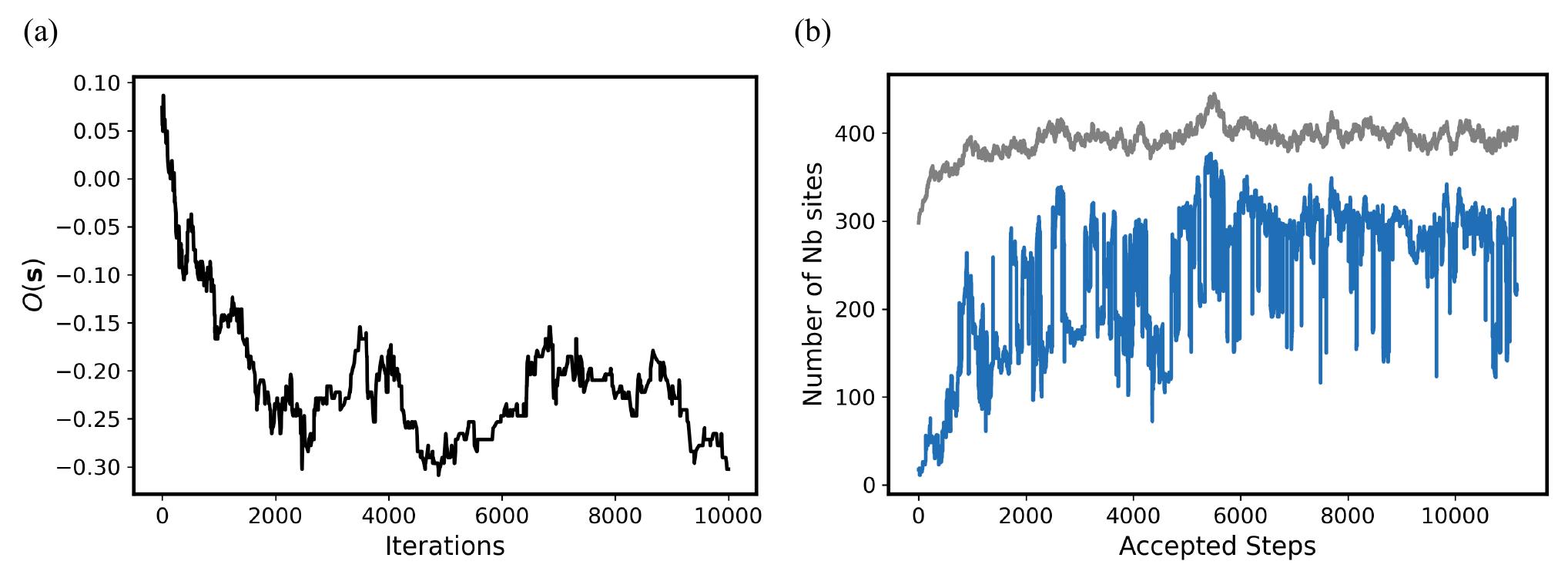}
  \caption{(a) Order parameter $O(s)$ as a function of FIRE-Swap iterations. The results are from DP-driven simulation of $6\times6\times6$ supercell with $T_{\mathrm{FS}}=1200$K. (b) The number of Nb sites in all Nb clusters (gray) and in the largest cluster (blue). The results are from DP-driven simulation of $12\times12\times12$ supercell with $T_{\mathrm{FS}}=1200$K. }
  \label{fig:1200K}

\end{figure}

\clearpage

\section{Evolution of cluster size in FIRE-Swap simulations of PMN, PZT and PST}

\begin{figure}[h]
  \centering
  \includegraphics[width=0.8\textwidth]{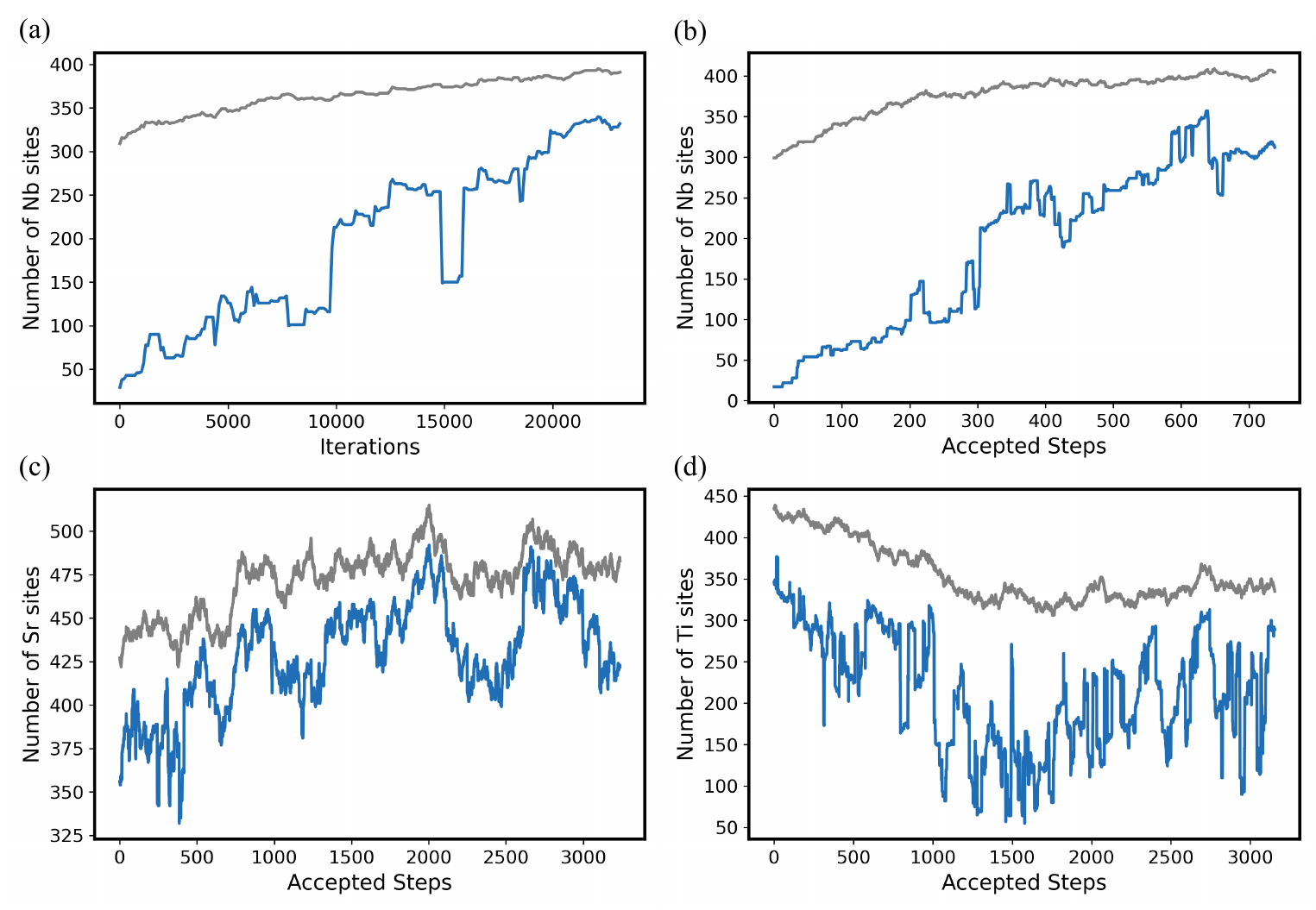}
  \caption{Evolution of cluster sizes during the FIRE-Swap simulations. (a) PMN with PBEsol-based CACE-LR. (b) PMN with UniPero. (c) PST with UniPero. (d) PZT with UniPero.}
  \label{sfig:cluster-size}

\end{figure}

Here, we report FIRE-Swap simulation results for $12\times12\times12$ supercells of PMN, PZT, and PST, as a supplement to the DP-driven results shown in Fig.~2(b) of the main text.
Supplemental Fig.~\ref{sfig:cluster-size} shows the number of majority B-site ions (Nb for PMN, Ti for PZT) or majority A-site ions (Sr for PST) in all clusters (gray) and in the largest cluster (blue). The four panels in Supplemental Fig.~\ref{sfig:cluster-size} correspond to (a) PMN with PBEsol-based CACE-LR, (b) PMN with UniPero, (c) PST with UniPero, and (d) PZT with UniPero. Across the DP, CACE-LR, and UniPero models, the largest Nb cluster in PMN always dominates the lattice and exhibits a surface-to-volume ratio of about 3. Meanwhile, Mg clusters are nearly absent because Mg--Mg nearest neighbors are strongly disfavored by short-range interactions, including electrostatic repulsion. 

Meanwhile, PZT and PST do not exhibit long-range chemical order. From the optimized compositional structures, we find an approximately homogeneous random distribution of small Zr clusters in PZT and small Pb clusters in PST. Comparing PST and PZT at the same mixing ratio, we also find that Pb clusters in PST are generally larger and more numerous than Zr clusters in PZT, due to differences in short-range interactions. Aggregation of the minority metallic ions violates the assumptions of standard percolation theory and simultaneously facilitates the percolation of the majority metallic ions. Consistently, Supplemental Fig.~\ref{sfig:cluster-size}(c) shows that the largest Sr cluster in PST also dominates the lattice. In PZT, by contrast, standard percolation assumptions approximately apply, and the largest Ti cluster remains relatively small compared to the size of the supercell.

\clearpage

\section{Finite size analysis}

In DP-driven FIRE-Swap simulations of $6\times6\times6$ PMN supercells, we often observe in the optimized compositional structures a finite-size artifact that does not break the Nb-filled sublattice or increase the order parameter $O$. Specifically, Nb sites form a slab-shaped cluster, leading approximately to a $(\mathrm{PbNbO_3})_2(\mathrm{PbNb_{1/2}Mg_{1/2}O_3})_4$ superlattice structure.
This artifact indicates phase separation between Nb-rich and rock-salt clusters, which can occur without changing the composition of sublattice $\beta^{\mathrm{II}}$. However, the supercell here ($N_c=216$) is too small to relax the interfaces between clusters.
Therefore, in the main text, we use the $12\times12\times12$ supercell to study clustering, which allows the cluster interfaces to relax more freely.

Here we provide evidence that supercells larger than $12\times12\times12$ produce quantitatively similar clustering results. In Supplemental Fig.~\ref{sfig:finite-size}, we compare DP-driven FIRE-Swap simulations of $12\times12\times12$ ($N_c=1728$) and $12\times12\times24$ ($N_c=3456$) PMN supercells, both performed at $T_{\text{FS}}=600$ K.

It is evident that both the total number of Nb sites in all Nb clusters and the number of Nb sites in the largest Nb cluster scale with supercell size. We also find that $O(s)$ remains close to $-1/3$ throughout both simulations. No spurious superlattice structures are observed in either case.

\begin{figure}[h]
  \centering
  \includegraphics[width=0.8\textwidth]{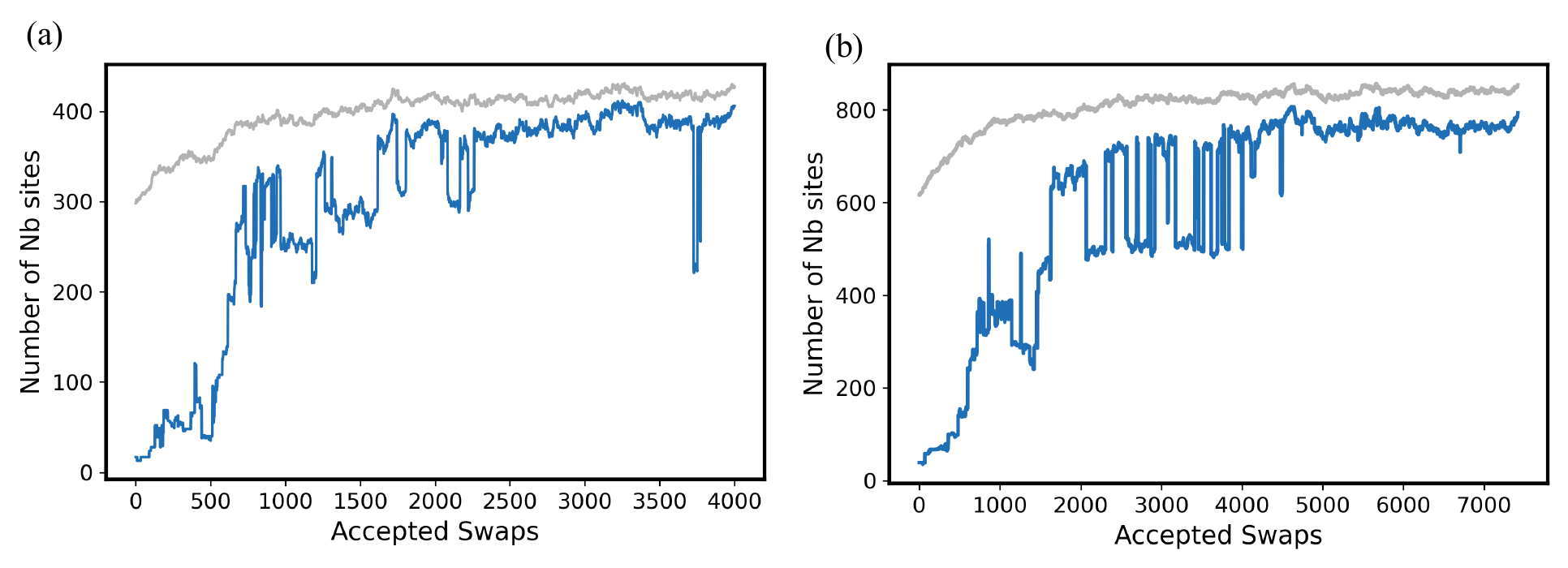}
  \caption{The number of Nb sites in all Nb clusters (gray) and in the largest cluster (blue). The results are from DP-driven simulation of $12\times12\times12$ (panel (a)) and $12\times12\times24$ (panel (b)) supercells for PMN. }
  \label{sfig:finite-size}

\end{figure}

\clearpage
\section{$H(r)$ and $g(r)$ associated to different compositional structure}

As stated in the main text, the mapping from scattering measurements to chemical order is subtle and non-unique. Here we provide evidence that the pair distribution function $H(r)$ and the radial distribution function $g(r)$ are not sensitive to changes in compositional structures.

For the initial and final compositional structures sampled in the DP-driven FIRE-Swap simulation ($T_{\text{FS}}=600$ K) of the $12\times12\times12$ PMN supercell (indicated in the upper part of Supplemental Fig.~\ref{sfig:rdf}), we perform room-temperature (300 K) NVT-MD simulations and calculate the corresponding $H(r)$ and $g(r)$ (lower part of Supplemental Fig.~\ref{sfig:rdf}). We find that, despite substantial changes in compositional structure, both $H(r)$ and $g(r)$ change only slightly.

These observations are consistent with the fact that $H(r)$ and $g(r)$ are averaged geometric quantities that integrate over many distinct local environments. In a complex perovskite, different metallic-ion arrangements can produce similar pair-distance distributions after thermal and configurational averaging, especially beyond the first coordination shell. Consequently, $H(r)$ and $g(r)$ are sensitive to average bond lengths and broadening mechanisms (e.g., thermal motion and model error), but they are much less sensitive to the topology and connectivity of chemical order that FIRE-Swap is designed to resolve.

\begin{figure}[h]
  \centering
  \includegraphics[width=0.9\textwidth]{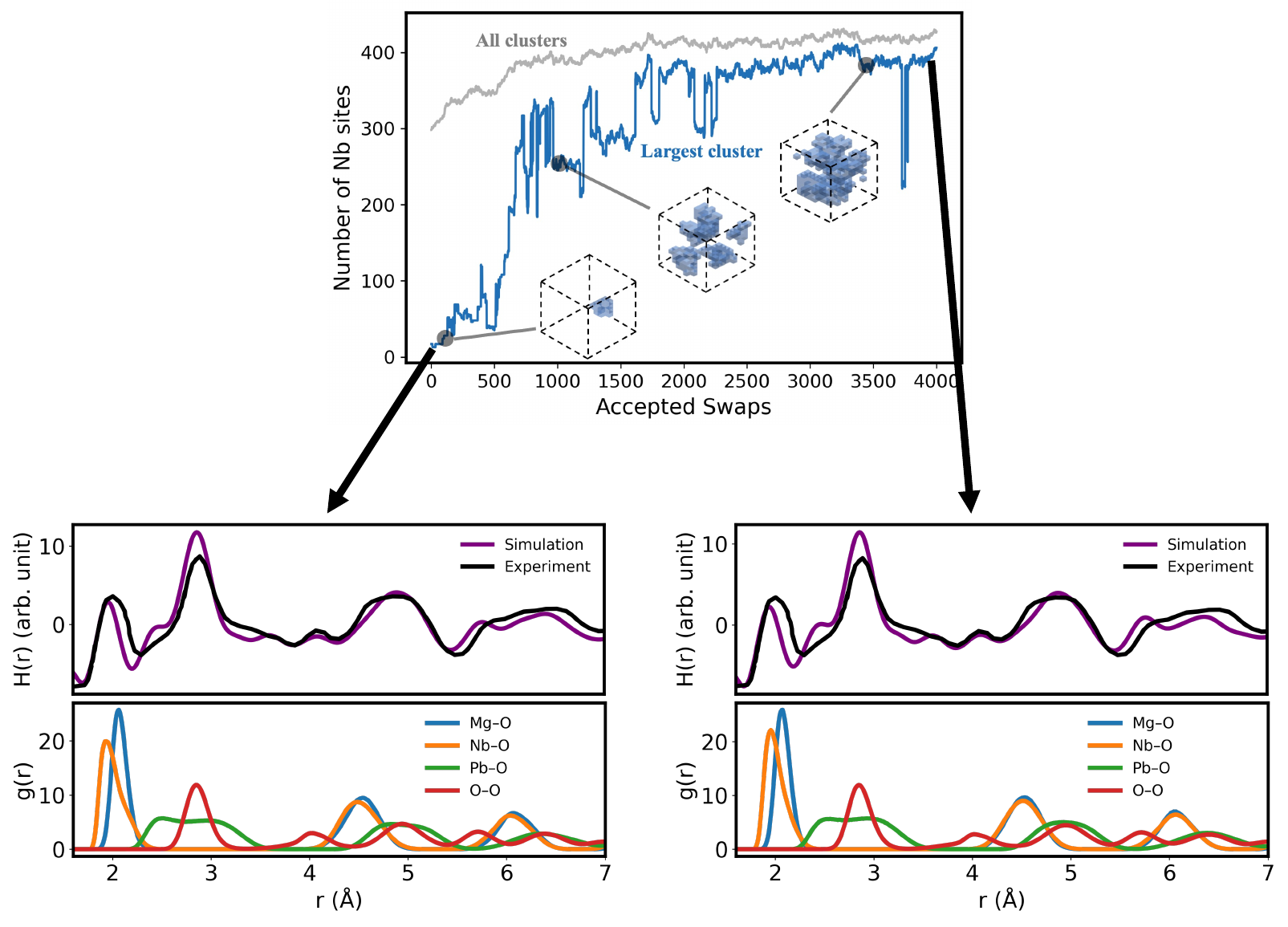}
  \caption{Upper part: indication of the sampled composition structures used for calculating $H(r)$ and $g(r)$. Lower part:
  The predicted (purple) and experimental (black) $H(r)$ associated with neutron scattering of PMN powder samples at $T=300$K. The predicted $g(r)$ are also reported. }
  \label{sfig:rdf}
\end{figure}

\clearpage
\section{Distribution of Pb-based local order parameter}

Here, we study a lead-based local order parameter, $\mathbf{d}_{\text{Pb}}$, that reflects local polarity in PMN. $\mathbf{d}_{\text{Pb}}$ is defined
as the displacement of a Pb ion away from the geometric center of its 12 surrounding oxygen ions.
The histogram of the probability density function $p(d_{\text{Pb}}^x, d_{\text{Pb}}^y)$ is reported in Supplemental Fig.~\ref{sfig:dipole-pdf}.
This histogram is extracted from the same set of molecular dynamics trajectories used to calculate the histograms in Fig.5(a) of the main text. So  these histograms can be directly compared.

From Supplemental Fig.~\ref{sfig:dipole-pdf}, the distribution of $\mathbf{d}_{\text{Pb}}$ displays similar qualitative behaviors as those reflected by the B-site-based local order parameter: quenched disorder manifests itself below Burn temperature, and the probability density distribution becomes less diffusive with emerging peaks when temperature decreases. But the distribution of $\mathbf{d}_{\text{Pb}}$ can not straightforwardly reveal the connection between compositional structure (clustering) and local polarity.

\begin{figure}[h]
  \centering
  \includegraphics[width=0.9\textwidth]{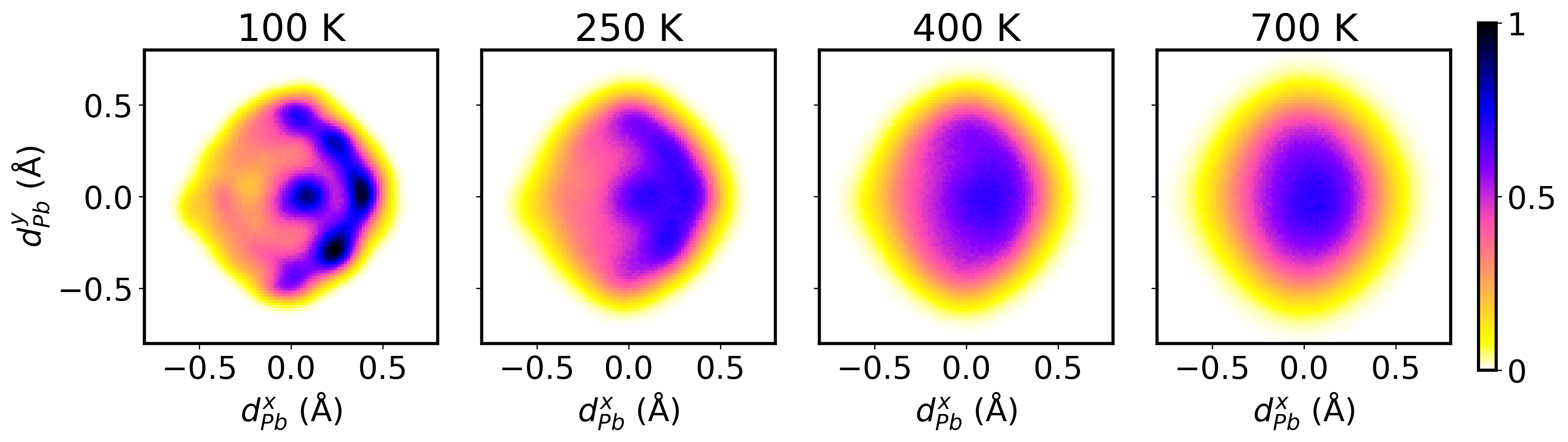}
  \caption{Histograms of PDF $p(d_{\text{Pb}}^x, d_{\text{Pb}}^y)$ associated with different thermal temperatures. }
  \label{sfig:dipole-pdf}
\end{figure}

\clearpage
\section{Error Distribution of Deep Potential model}

We report the error distribution of the productive Deep Potential model relative to the SCAN-DFT dataset in Supplemental Fig.~\ref{Fig:dperror}.

\begin{figure}[htb]
  \centering
  \includegraphics[width=0.9\linewidth]{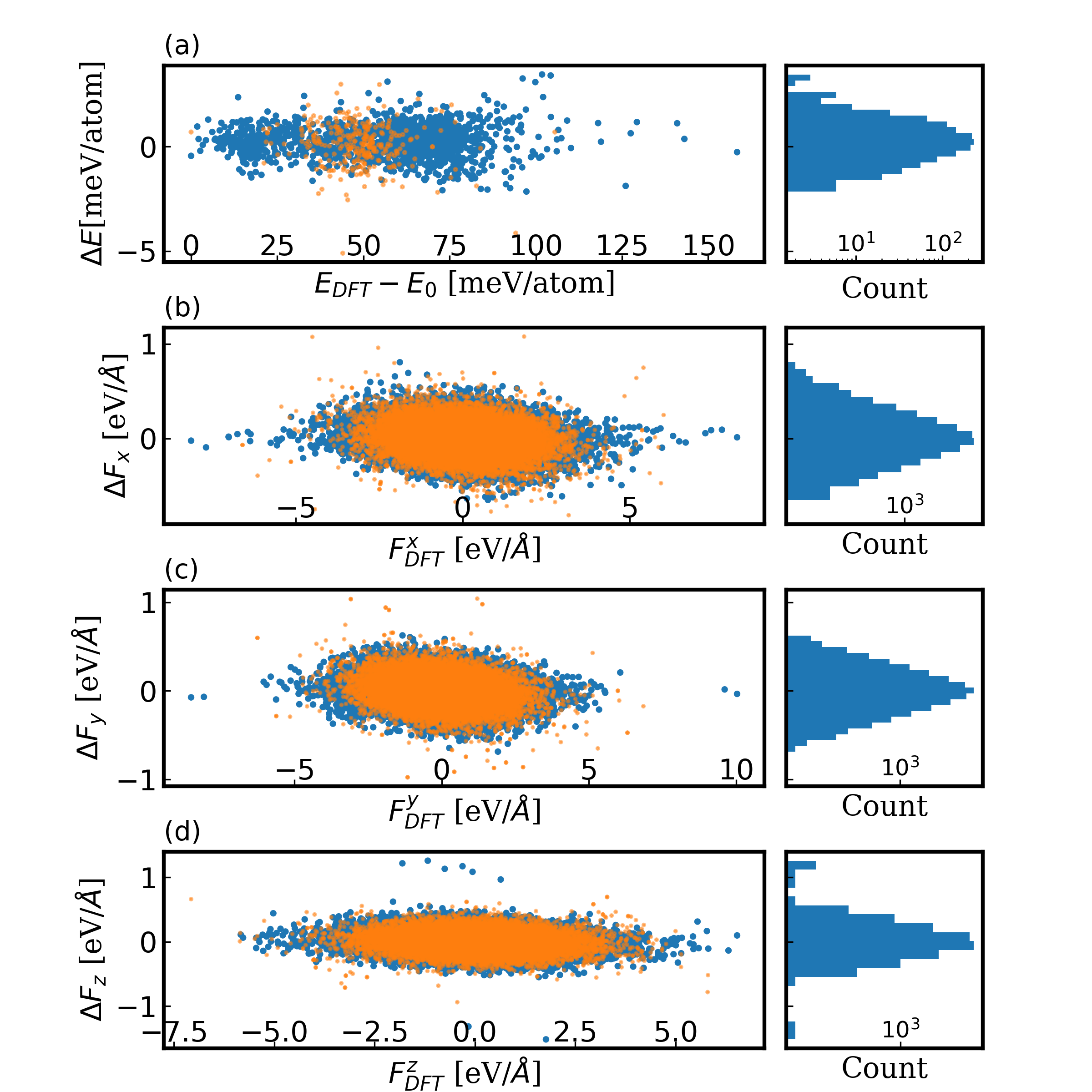}
  \caption{Error distribution of the DP model on the train set (blue) and the test set (orange).}
  \label{Fig:dperror}
\end{figure}
